\documentclass[aps,prd,preprint,groupedaddress,showpacs]{revtex4}

\usepackage{epsfig}
\usepackage{graphics}

\begin{document}

\leftmargin -2cm
\def\choosen{\atopwithdelims..}

\boldmath
\title{Deep inelastic scattering and prompt photon production within the framework of quark Reggeization hypothesis}
\unboldmath
  \author{\firstname{V.A.} \surname{Saleev}}
\email{saleev@ssu.samara.ru}

\affiliation{ Samara State University, Ac. Pavlov St. 1, 443011
Samara, Russia}

\begin{abstract}
We study  $ep$ deep inelastic scattering  and the inclusive
production of prompt photon within the framework of the
quasi-multi-Regge-kinematic approach, applying the quark
Reggeization hypothesis. We describe structure functions $F_2$ and
$F_L$ supposing that a virtual photon scatters on a Reggeized quark
from a proton, via the effective gamma-Reggeon-quark vertex. It is
shown that the main mechanism of the inclusive prompt photon
production in  $p\bar p$ collisions is the fusion of a Reggeized
quark and a Reggeized antiquark into a photon, via the effective
Regeon-Reggeon-gamma vertex.  We describe the inclusive photon
transverse momentum  spectra measured by the CDF and D0
Collaborations within errors and without free parameters, using the
Kimber-Martin-Ryskin unintegrated quark and gluon distribution
functions in a proton.
\end{abstract}


\maketitle \maketitle

\section{Introduction}
\label{sec:one}

It is well known, that studies of lepton deep inelastic scattering
(DIS) and production of photons with large transverse momenta
producing in the hard interaction between two partons in hadron
collisions, so called prompt photon production, provide precision
tests of perturbative quantum chromodynamics (QCD) as well as
information on the parton densities within protons.

Also these studies are  our potential for the observation of a new
dynamical regime, namely the high-energy Regge limit, which is
characterized by the following condition
$\sqrt{S}>>\mu>>\Lambda_{QCD}$, where $\sqrt{S}$ is the total
collision energy in the center of mass reference frame,
$\Lambda_{QCD}$ is the asymptotic scale parameter of QCD, $\mu$ is
the typical energy scale of the hard interaction. At this high
energy limit, the contribution from the partonic subprocesses
involving $t-$channel parton (quark or gluon) exchanges to the
production cross section can become dominant. In the region under
consideration, the transverse momenta of the incoming partons and
their off-shell properties can no longer be neglected, and we deal
with "Reggeized" $t-$channel partons.

The theoretical frameworks for this kind of high-energy
phenomenology are the $k_T-$factorization approach
\cite{GribovLevinRyskin,CollinsEllis,CataniCH} and the
quasi-multi-Regge kinematics (QMRK) approach \cite{FadinLipatov96}.
Our previous analysis of charmonium and bottomonium production at
the Fermilab Tevatron and DESY HERA Colliders using the high-energy
factorization scheme \cite{KniehlSaleevVasin} has shown the
appreciation of the method even at the leading order (LO) in the
strong-coupling constant $\alpha_s$ in comparison  with the
experimental data. We have found that essential features produced by
the high-energy factorization scheme at LO are being at
next-to-leading (NLO) in the conventional collinear parton model.

On the contrary the $k_T-$factorization scheme, based on the well
known prescription for off-shell $t-$channel gluons
\cite{CollinsEllis}, the QMRK approach seems to be more proper
theoretically \cite{Lipatov95}. In this approach we work with gauge
invariant amplitudes and use the factorization hypothesis, which is
proved in the leading-logarithmic-approximation (LLA). Recently it
was shown, that the calculation at the NLO in the strong coupling
constant within the framework of the QMRK approach can be done
\cite{NLO}.

In this paper we study the $ep$ DIS and the inclusive production of
prompt photon within the framework of the QMRK approach, applying
the quark Reggeization hypothesis \cite{FadinSherman}. We describe
DIS structure functions $F_2$ and $F_L$ suggesting that a virtual
photon scatters on a Reggeized quarks from a proton, via the
effective gamma-Reggeon-quark vertices. It is shown that the main
mechanism of the inclusive prompt photon production in  $p\bar p$
collisions is the fusion of a Reggeized quark and a Reggeized
antiquark into a photon, via the effective Reggeon-Reggeon-gamma
vertices. We describe the inclusive photon transverse momentum and
pseudorapidity spectra measured by the D0\cite{D018,D0196}, and
CDF\cite{CDF18} Collaborations within errors and without free
parameters, using Kimber-Martin-Ryskin unintegrated \cite{KMR} quark
and gluon distribution functions in a proton. In case of prompt
photon production we compare our results with  another ones obtained
recently in the $k_T-$factorization approach \cite{KMRgamma,
ZotovLipatovGamma, SzczurekGamma},  in detail.

 This paper is organized as follows.
In Sec.~\ref{sec:two}, the relevant Reggeon-Reggeon-Particle (RRP)
and Particle-Reggeon-Particle (PRP) effective vertices are written
and discussed. In Sec.~\ref{sec:three}, we consider $ep$ DIS in the
QMRK approach, applying the quark Reggeization hypothesis.  In
Sec.~\ref{sec:foure}, we describe in the high-energy factorization
scheme the inclusive photon production at the Tevatron Collider both
directly via the effective  Reggeon-Reggeon-gamma vertices, and in
the quark fragmentation into a photon via the
Reggeon-Reggeon-Particle vertex.  In Sec.~\ref{sec:five}, we discuss
and compare the relevant approaches used in the another studies
within the $k_T-$factorization approach. In Sec.~\ref{sec:six}, we
summarize our conclusions.

%\boldmath
\section{Basic Formalism}
%\unboldmath
\label{sec:two}

In the phenomenology of strong interactions at high energies, it is
necessary to describe the QCD evolution of the parton distribution
functions of  colliding particles starting with some scale $\mu_0$,
which controls a non-perturbative regime, to the typical scale $\mu$
of the hard-scattering processes, which is typically of the order of
the transverse mass $M_T=\sqrt{M^2+|{\bf p}_T|^2}$ of the produced
particle with mass $M$ and transverse momentum ${\bf p}_T$. In the
region of very high energies, in so-called Regge limit, the typical
ratio $x=\mu/\sqrt{S}$ becomes very small, $x\ll1$. That leads to
large logarithmic contributions of the type $[\alpha_s\ln(1/x)]^n$
in the resummation procedure, which is described by the BFKL
evolution equation \cite{BFKL} for an unintegrated gluon (quark)
distribution function $\Phi_{g,q}(x,|{\bf q}_T|^2,\mu^2)$.
Correspondingly, in the QMRK approach \cite{FadinLipatov96} the
initial-state $t$-channel gluons and quarks are considered as
Reggeons, or  Reggeized gluons $(R)$ and Reggeized quarks $(Q)$.
They are off-mass shell and carry finite transverse two-momenta
${\bf q}_T$ with respect to the hadron beam from which they stem.

Recently, in the Ref.~\cite{KTAntonov}, the Feynman rules for the
effective theory based on the non-abelian gauge-invariant action
\cite{Lipatov95} were derived for the induced and some important
effective vertices. However, these rules include  processes with
Reggeized gluons in the initial state only. In case of $t$-channel
quark exchange processes such rules are unknown today and it is
necessary to construct effective vertices with the Reggeized quarks,
using QMRK approach prescriptions, every time from the begining. Of
course, a certain set of Reggeon-Reggeon-Particle effective vertices
are known, for example for the transitions $R R\to g$
\cite{LipatovFadin89} and $Q\bar Q\to g$  \cite{FadinSherman}.

Roughly speaking, the Reggeization of amplitudes is a trick, which
gives an opportunity to take into account efficiently large
radiative corrections to the processes under Regge limit condition
beyond the collinear approximation. The particle Reggeization is
known in high energy quantum electrodynamics (QED) for electrons
only \cite{GellMann} and for gluons and quarks in QCD
\cite{BFKL,FadinSherman}.

Contrary to the usual QCD vertices such as, for example,
quark-quark-gluon vertex for which we can draw a definite set of
Feynman diagrams with perfectly defined rules for the calculation of
their contributions, we have no similar rules for the Reggeized
quark vertices. These vertices are extracted from the comparison of
radiative corrections to the scattering amplitudes with their
Reggeized forms \cite{FadinFiore}. Details of a reduction procedure
from the initial set of diagrams with the on-shell collinear quarks
or gluons to one effective Reggeon-Reggeon-Particle (or
Particle-Reggeon-Particle) vertex can be found, for example, in the
following papers \cite{FadinFiore, LipatovFadin89}. We will use this
method for the gamma-Reggeon-quark $(\gamma^\star Q\to q)$ vertex,
which will be used in our calculations below.

\begin{figure}[ht]
\begin{center}
\includegraphics[width=1.0\textwidth, clip=]{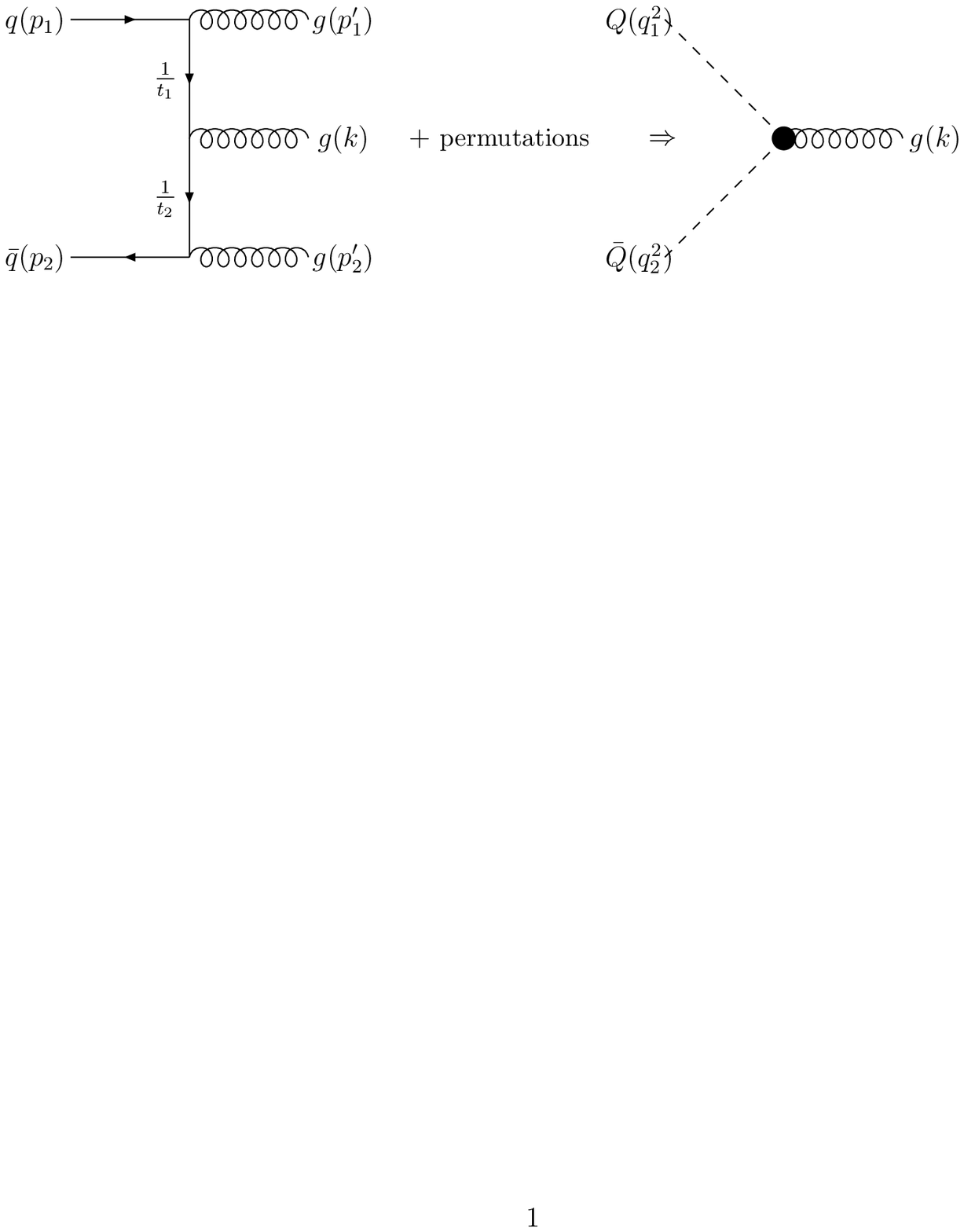}
\end{center}
\caption{The $C_{Q\bar Q}^g$ effective vertices. \label{fig:qq}}
\end{figure}

Thus, the effective Reggeon-Reggeon-gluon (or
Reggeon-Reggeon-photon) vertex $C_{\bar Q Q}^{g(\gamma)}(q_1,q_2)$
can be extracted from any amplitude of the $g(\gamma)$ production in
the QMRK approach. In the simple case, they can be extracted from
the amplitude $A_{\bar q q}^{g(g)g}$, which describes gluon (photon)
production in the process:
\begin{equation}
q(p_1) + \bar q(p_2) \to g(p_1') + g(k) + g(p_2')\label{eq:qqGgg}.
\end{equation}
At the high energy limit this process is described by a set of
ladder type Feynman quark-exchange diagrams, which are presented in
the Fig.~\ref{fig:qq}. For the external gluon we use  physical
polarizations with different gauge fixing conditions for gluons
moving along $p_1$ and $p_2$ , thus if the gluon momentum $p_1'$ has
a large component along $p_1$, its polarization vector $e(p_1')$
satisfies equations $e(p_1')p_1'= e(p_1')p_2=0$. The Reggeized form
of the amplitude is expressed as follows:
\begin{equation}
A_{\bar q q}^{g(g)g}=2s\Gamma_{\bar q g}^{\bar Q}
\frac{1}{t_1}C_{\bar Q Q}^g \frac{1}{t_2}\Gamma_{q g}^Q
\label{eq:qmrkamp},
\end{equation}
where $t_1=(p_1-p_1')^2$, $t_2=(p_2-p_2')^2$, $s=2(p_1p_2)$,
$\Gamma_{\bar q g}^{\bar Q}$ and  $\Gamma_{q g}^Q$ are the LO
Particle-Particle-Reggeon vertices, see Ref.~\cite{FadinFiore}. The
effective vertex $C_{\bar Q Q}^g(q_1,q_2)$, which describes the
production of a single gluon with the momentum $k(g)=q_1(Q)+q_2(\bar
Q)$ in the Reggeized quark and Reggeized antiquark  annihilation was
obtained in Ref.~\cite{FadinSherman,FadinBogdan} and it can be
presented, omitting the color and Lorentz indices  in the left side,
as follows:
\begin{equation}
C_{\bar Q Q}^g(q_1,q_2)=D_a^\mu(q_1,q_2)=-g_s T^a \left( \gamma^\mu
-\frac{2P_1^\mu }{x_2S}\hat{q_2}-\frac{2P_2^\mu
}{x_1S}\hat{q_1}\right)\label{eq:QQg},
\end{equation}
where $P_{1,2}$ are colliding hadron momenta,
$P_{(1,2)}=E_{(1,2)}(1,\vec 0,\pm 1)$ and $S=2(P_1P_2)=4E_1E_2$. The
Reggeized quark momenta are $q_1=x_1P_1+q_{1T}$,
$q_2=x_2P_2+q_{2T}$, and $q_{(1,2)T}=(0,\vec q_{(1,2)T},0)$. It is
obviously that this vertex satisfies the gauge invariant condition
$D_{a}^\mu(q_1,q_2) k_\mu=0$.

The same as for the gluon production, the effective vertex for the
photon production can be written in the form:
\begin{equation}
C_{\bar Q Q}^\gamma(q_1,q_2)=F^\mu(q_1,q_2)=-e e_q \left( \gamma^\mu
-\frac{2P_1^\mu }{x_2S}\hat{q_2}-\frac{2P_2^\mu
}{x_1S}\hat{q_1}\right)\label{eq:QQgamma},
\end{equation}
where $e=\sqrt{4\pi\alpha}$ is the electromagnetic coupling
constant, $e_q$ is the quark electric charge.

The effective vertex $C_{RR}^g(q_1,q_2)$, which describes the
production of a single gluon in the collision of two Reggeized
gluons has the following form \cite{LipatovFadin89}:
\begin{equation}
C_{RR}^g(q_1,q_2)=W_{abc}^\mu(q_1,q_2)=-g_s
f_{abc}\left((q_2-q_1)^\mu+2P_1^\mu(x_1+
\frac{q_1^2}{x_2S})-2P_2^\mu(x_2+\frac{q_2^2}{x_1S})\right)\label{eq:RRg}.
\end{equation}

\begin{figure}[ht]
\begin{center}
\includegraphics[width=1.0\textwidth, clip=]{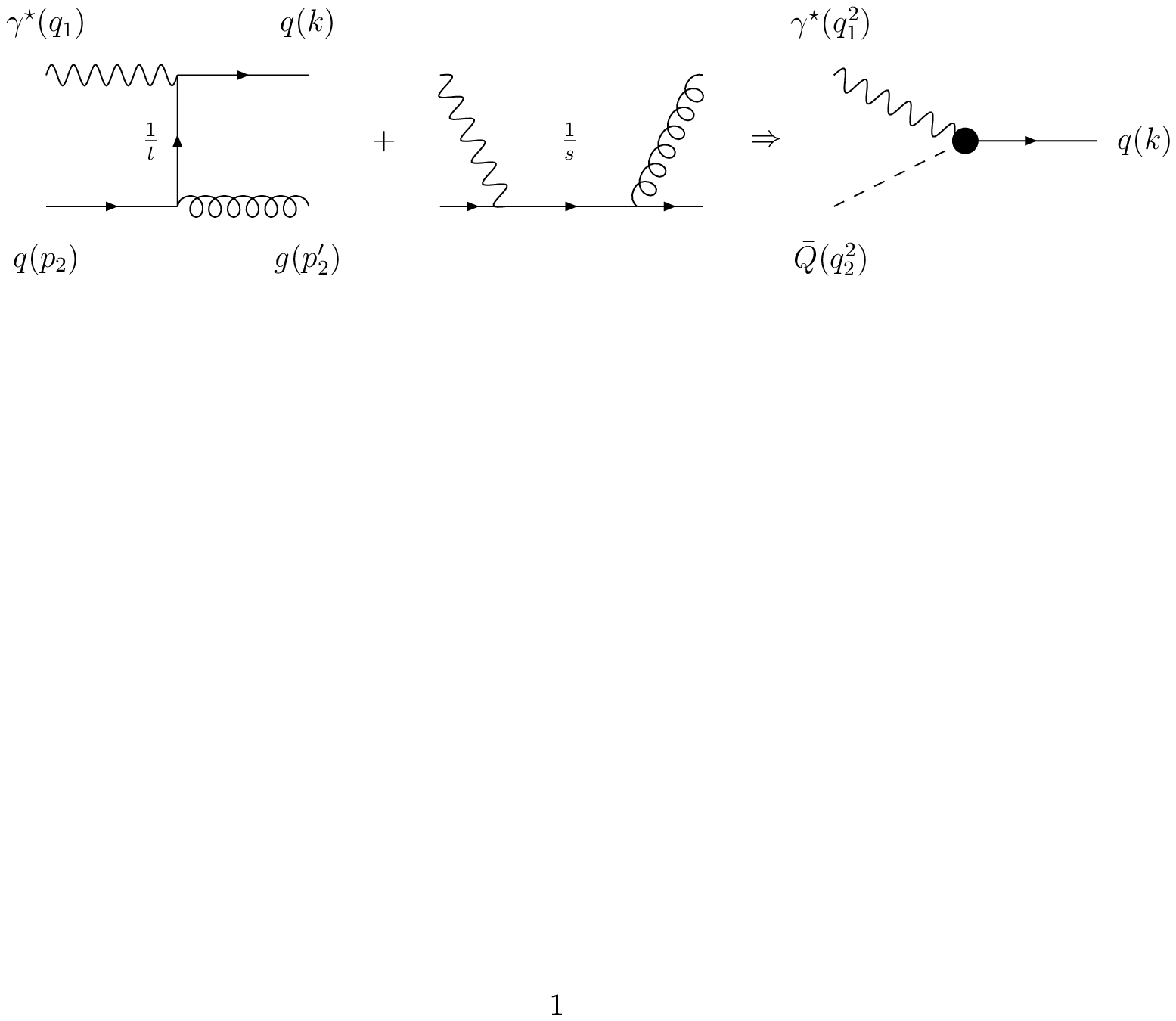}
\end{center}
\caption{The $C_{\gamma Q}^q$ effective vertices.
\label{fig:gammaq}}
\end{figure}

The effective vertex, which describes the production of a quark with
the momentum $k=q_1(\gamma^\star)+q_2(Q)$ in a virtual photon
collision with a Reggeized quark can be extracted from the
amplitude, which is described by the two diagrams presented in
Fig.~\ref{fig:gammaq}, and it is written as follows
\begin{equation}
C_{\gamma Q}^q(q_1,q_2)=G^\mu(q_1,q_2)=-e e_q\left(
\frac{q_1^2}{q_1^2+q_2^2}\gamma^\mu-\frac{2k^\mu}{q_1^2+q_2^2}\hat{q_2}
+\frac{2q_2^2 x_2
P_2^\mu}{(q_1^2+q_2^2)^2}\hat{q_2}\right)\label{eq:gammaQq},
\end{equation}
where $q_2=x_2P_2+q_{2T}$ is the Reggeized quark four-momentum and
the gauge invariance condition is given by $G^\mu(q_1,q_2)q_{1\mu}
=0.$

The effective Reggeon-Reggeon-Particle vertex, which describes the
production of a quark with the momentum $k(q)=q_1(Q)+q_2(R)$ in the
Reggeized quark  and Reggeized gluon collision can be presented as
follows
\begin{equation}
C_{R Q}^q(q_1,q_2)=V(q_1,q_2)=-g_s T^a \hat \varepsilon_T
(q_2)\label{eq:RQq},
\end{equation}
where $\hat \varepsilon_T(q_2)=\gamma_\mu \varepsilon^\mu(q_2)$ and
$\varepsilon^\mu(q_2)=\displaystyle{\frac{q_{2T}^\mu}{|\vec q_{2T}
|}}$. Let us note, that contrary to the definition used in the
Ref.~\cite{FadinBogdan} in the relevant cases, we don't include
quark spinors  in the our formulas for the effective vertices
(\ref{eq:QQg})-(\ref{eq:RQq}).

\boldmath
\section{Deep inelastic scattering and quark Reggeization}
\unboldmath \label{sec:three}

In this part we consider electron (positron)  DIS on a proton at
high energy in the framework of the quark Reggeization hypothesis.
At first, to write down the conventional kinematic DIS formulas we
define relevant four-momenta: $P_e$ is the electron initial
four-momentum, $P_N$ is the proton four-momentum, $q_1$ is the
virtual photon four-momentum, $q_2$ is the Reggeized quark
four-momentum. The invariant variables are defined as follows:
\begin{equation}
S_{eN}=2(P_eP_N), \quad y=\displaystyle{\frac{(q_1P_N)}{(P_eP_N)}},
\quad x_B=\frac{Q^2}{2(q_1P_N)}, \quad Q^2=-q_1^2.
\end{equation}

As usual, we write the differential cross section as a convolution
of the lepton tensor $L_{\mu\nu}$ and the hadron tensor $W^{\mu\nu}$
\begin{equation}
\frac{d\sigma(eN)}{dx_BdQ^2}=\frac{2\pi\alpha^2}{x_BQ^4}\frac{m_N
y}{S_{eN}}L_{\mu\nu}W^{\mu\nu},
\end{equation}
where $m_N$ is the proton mass. The last one can be presented in
terms of DIS structure functions $F_2(x_B,Q^2)$ and $F_L(x_B,Q^2)$
\begin{equation}
\frac{d\sigma(eN)}{dx_BdQ^2}=\frac{2\pi\alpha^2}{x_BQ^4}\left[
(2-2y+y^2)F_2(x_B,Q^2)-y^2F_L(x_B,Q^2) \right].
\end{equation}
The structure functions $F_2(x_B,Q^2)$ and $F_L(x_B,Q^2)$ are
obtained by projection of the hadron tensor $W^{\mu\nu}$ as follows:
\begin{eqnarray}
F_2(x_B,Q^2)&=&x_Bm_N\left(-g_{\mu\nu}+\frac{12x_B^2}{Q^2}P_{N\mu}P_{N\nu}\right)W^{\mu\nu}\label{eq:f2}\\
F_L(x_B,Q^2)&=&\frac{8x_B^3}{Q^2}m_N
P_{N\mu}P_{N\nu}W^{\mu\nu}\label{eq:fl}.
\end{eqnarray}

In the high-energy factorization scheme the DIS cross section is
presented as a convolution of the unintegrated parton structure
function and the off-shell partonic cross section or  the so-called
impact factor \cite{CollinsEllis}. It is well known that in the
collinear approximation in the region of small $x_B$ the gluon
contribution via $e + g\to e+ q+\bar q$ subprocess into $e N$ cross
section dominates over the direct photon-quark scattering in the
subprocess $e + q\to e + q$. In the QMRK approach we deal with the
electron-Reggeon cross sections $\sigma(e Q)$ and $\sigma(e R)$ and
the relevant LO subprocesses are the following
\begin{eqnarray}
&&e + Q \to e + q,\label{eq:eQeq}\\
&&e + R \to e + q + \bar q.\label{eq:eReqq}
\end{eqnarray}
Our analysis shows the dominant role of the LO electron -- Reggeized
quark scattering in the electron--proton DIS. As we will see below
the experimental data for DIS structure functions $F_2$ and $F_L$
are described via the electron -- Reggeized quark scattering
(\ref{eq:eQeq}) and the contribution of the NLO subprocess
(\ref{eq:eReqq}) should  be small.

So, accordingly the high-energy factorization scheme the electron --
proton and the electron -- Reggeized quark cross sections are
connected as follows:
\begin{equation}
d\sigma(eN\to eX)=\sum_{q,\bar q}\int \frac{dx_2}{x_2}\int \frac{d^2
q_{2T}}{\pi}\Phi_{q,\bar q}(x_2,t_2,\mu^2) d\hat \sigma(e Q\to
eq),\label{eq:feN}
\end{equation}
where $\Phi_{q}(x_2,t_2,\mu^2)$ is the quark unintegrated
distribution function, $t_2=-q_2^2={\bf q}_{2T}^2$. Working in the
$\gamma^\star p$ center of mass reference frame and taking into
account the defined above effective vertex (\ref{eq:gammaQq}) and
the factorization formula (\ref{eq:feN}), we obtain the hadron
tensor in the form:
\begin{equation}
W^{\mu\nu}=\sum_{q,\bar q}
\frac{e_q^2x_B}{4m_NQ^2}\int\frac{dt_2}{x_2}\Phi_q(x_2,t_2,\mu^2)\mbox{Tr}\left[
\hat kG^\mu(q_1,q_2)\hat P_2 G^\nu(q_1,q_2)\right],
\end{equation}
where $e_q$ is the quark electric charge, and
$$x_2=x_B\left(\frac{Q^2+t_2}{Q^2}\right).$$ Making the projection of
the hadron tensor  $W^{\mu\nu}$ on DIS structure functions we obtain
the following answers, which can be compared with the experimental
data:

\begin{eqnarray}
F_2(x_B,Q^2)&=&\sum_{q,\bar q}e_q^2\int
dt_2\left(\frac{x_B}{x_2}\right)^3
\Phi_{q,\bar q}(x_2,t_2,\mu^2)\left(\frac{Q^4+6t_2Q^2+2t_2^2}{Q^4}\right),\label{eq:f2my}\\
F_L(x_B,Q^2)&=&\sum_{q,\bar q}e_q^2\int
dt_2\left(\frac{x_B}{x_2}\right)^3 \Phi_{q,\bar
q}(x_2,t_2,\mu^2)\left(\frac{4t_2}{Q^2}\right)\label{eq:flmy}.
\end{eqnarray}
In the collinear limit, $t_2\to 0$, we immediately obtain the well
known normalization relations:
\begin{eqnarray}
F_2(x_B,Q^2)&=&\sum_{q,\bar q}e_q^2\int dt_2 \Phi_{q,\bar
q}(x_B,t_2,\mu^2)=\sum_{q,\bar q}e_q^2 x_B f_{q,\bar q}(x_B,\mu^2)\\
F_L(x_B,Q^2)&=&0,
\end{eqnarray}
where $f_{q}(x_B,\mu^2)$ is the collinear quark distribution
function.

At the stage of numerical calculations we use the
Kimber-Martin-Ryskin (KMR) \cite{KMR} prescription for gluon and
quark unintegrated distribution functions. In fact, we use
unintegrated distribution functions which are presented by G.~Watt
as ${\bf C}$++ codes \cite{WattMartinRyskin}, see
http://www.hep.ucl.ac.uk/~watt/. We see that the KMR approach of
obtaining unintegrated distribution functions is intuitively close
to the method of obtaining effective vertices in the QMRK approach.
So, here we suggest that KMR unintegrated distribution functions and
squared amplitudes, obtained in the QMRK approach, can be used
together to calculate observed cross sections.

\begin{figure}[ht]
\begin{center}
\includegraphics[width=0.45\textwidth, clip=]{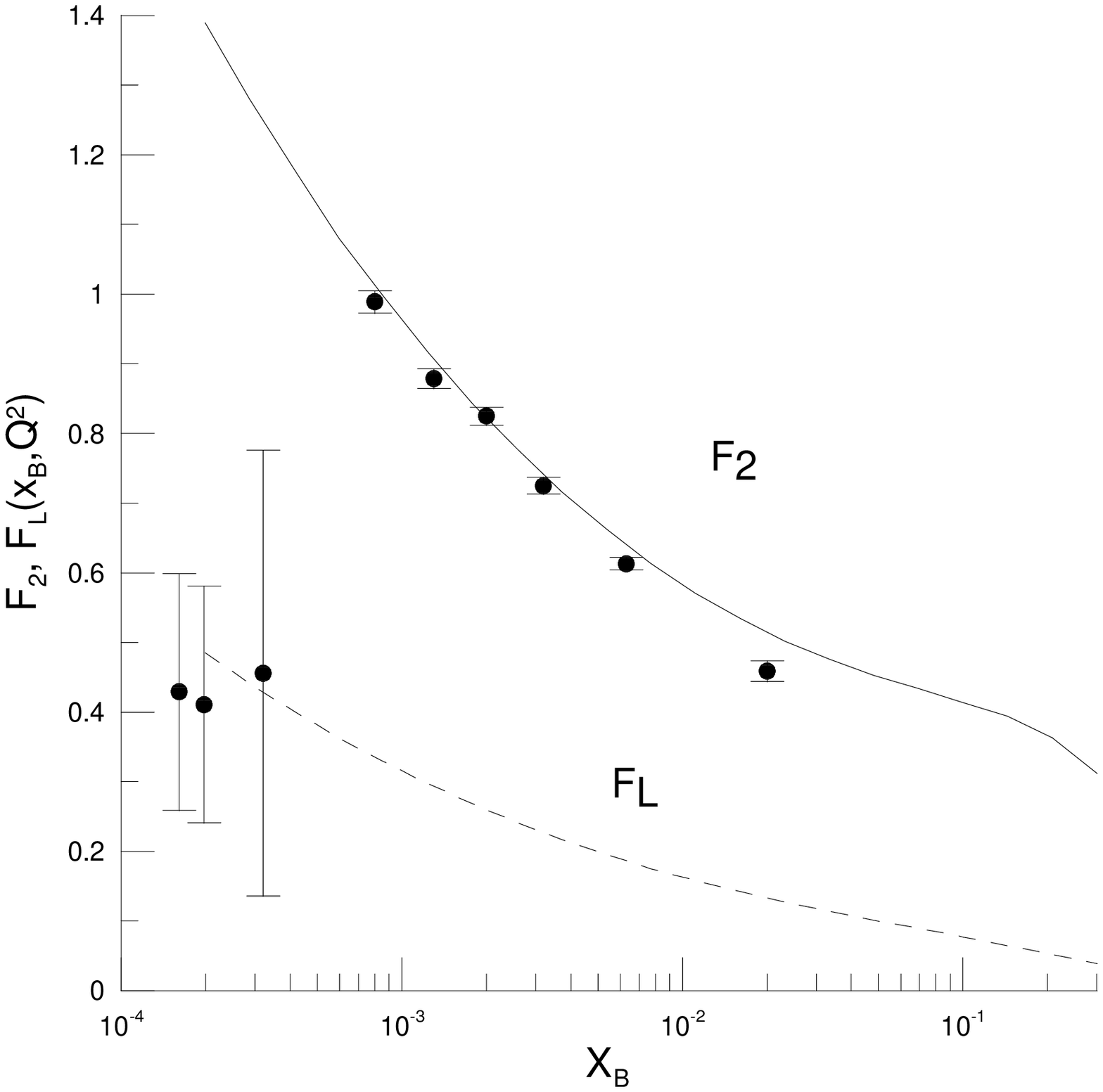}
\end{center}
\caption{$F_2(x_B,Q^2)$ and $F_L(x_B,Q^2)$ at $Q^2=12$ GeV$^2$. The
data are from H1 Collaboration \cite{H1DIS}. \label{fig:f12}}
\end{figure}

\begin{figure}[ht]
\begin{center}
\includegraphics[width=0.45\textwidth, clip=]{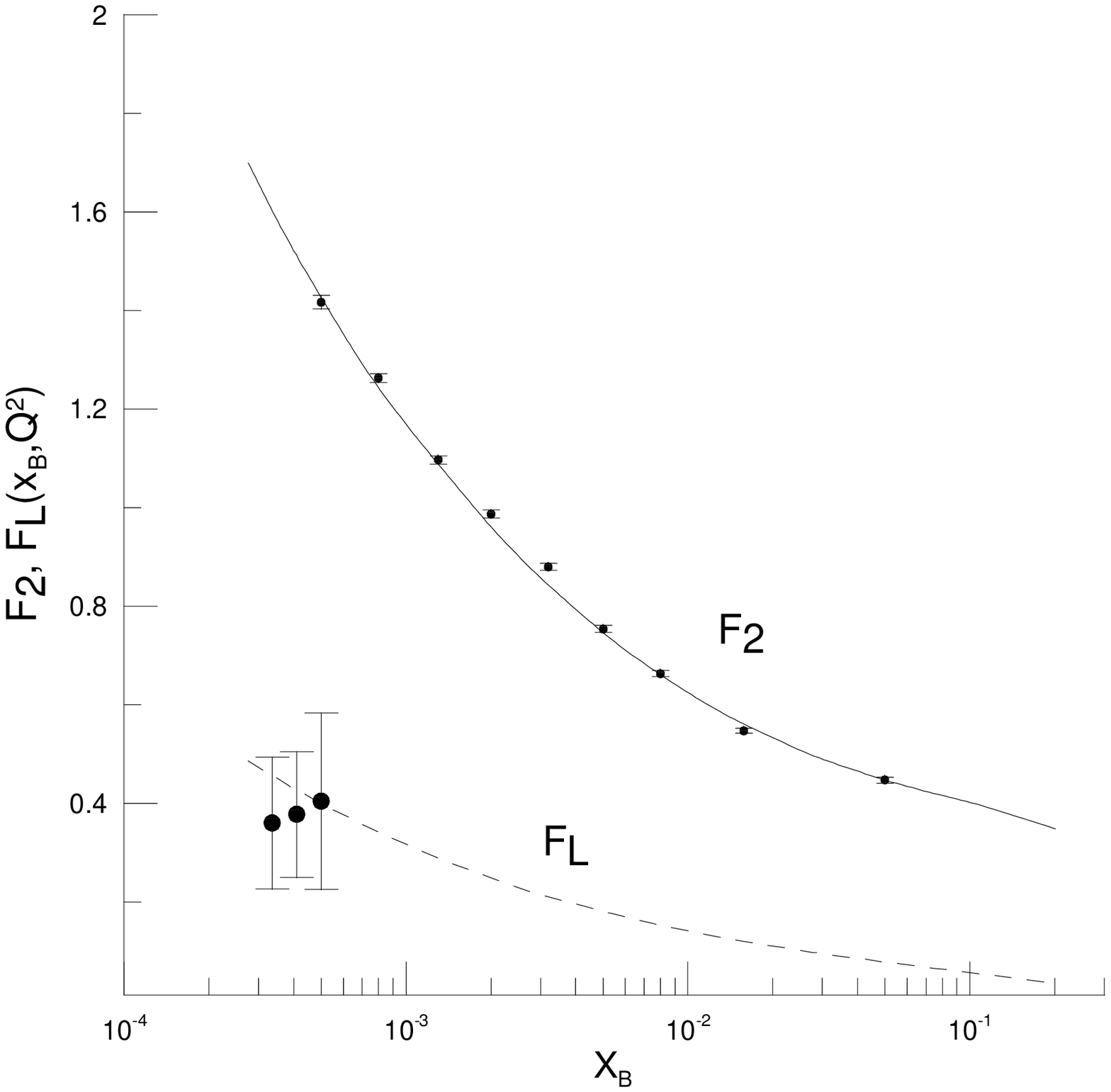}
\end{center}
\caption{$F_2(x_B,Q^2)$ and $F_L(x_B,Q^2)$ at $Q^2=25$ GeV$^2$. The
data are from H1 Collaboration \cite{H1DIS}. \label{fig:f25}}
\end{figure}

\begin{figure}[ht]
\begin{center}
\includegraphics[width=0.45\textwidth, clip=]{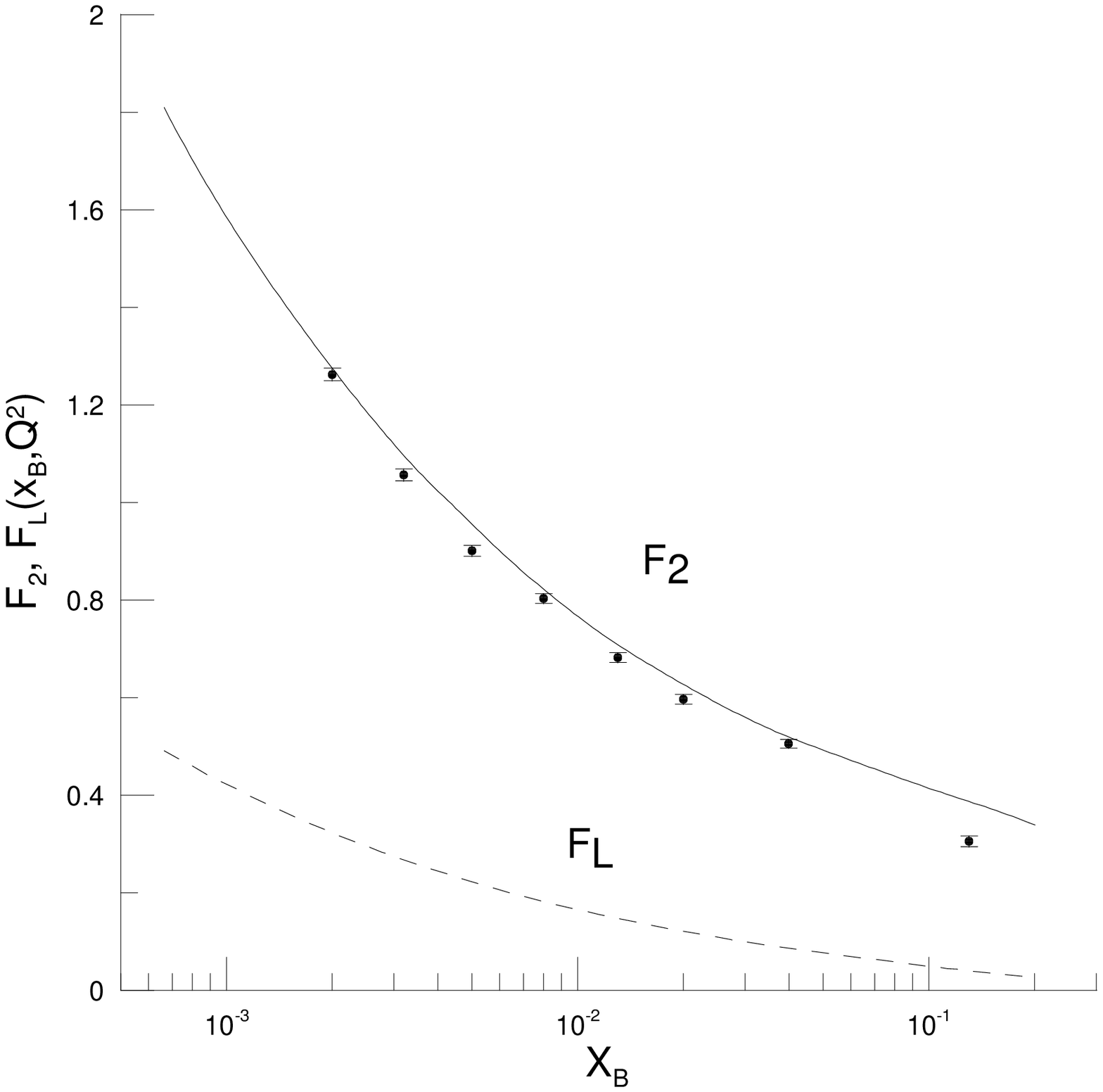}
\end{center}
\caption{$F_2(x_B,Q^2)$ and $F_L(x_B,Q^2)$ at $Q^2=60$ GeV$^2$ .The
data are from H1 Collaboration \cite{H1DIS}. \label{fig:f60}}
\end{figure}

\begin{figure}[ht]
\begin{center}
\includegraphics[width=0.45\textwidth, clip=]{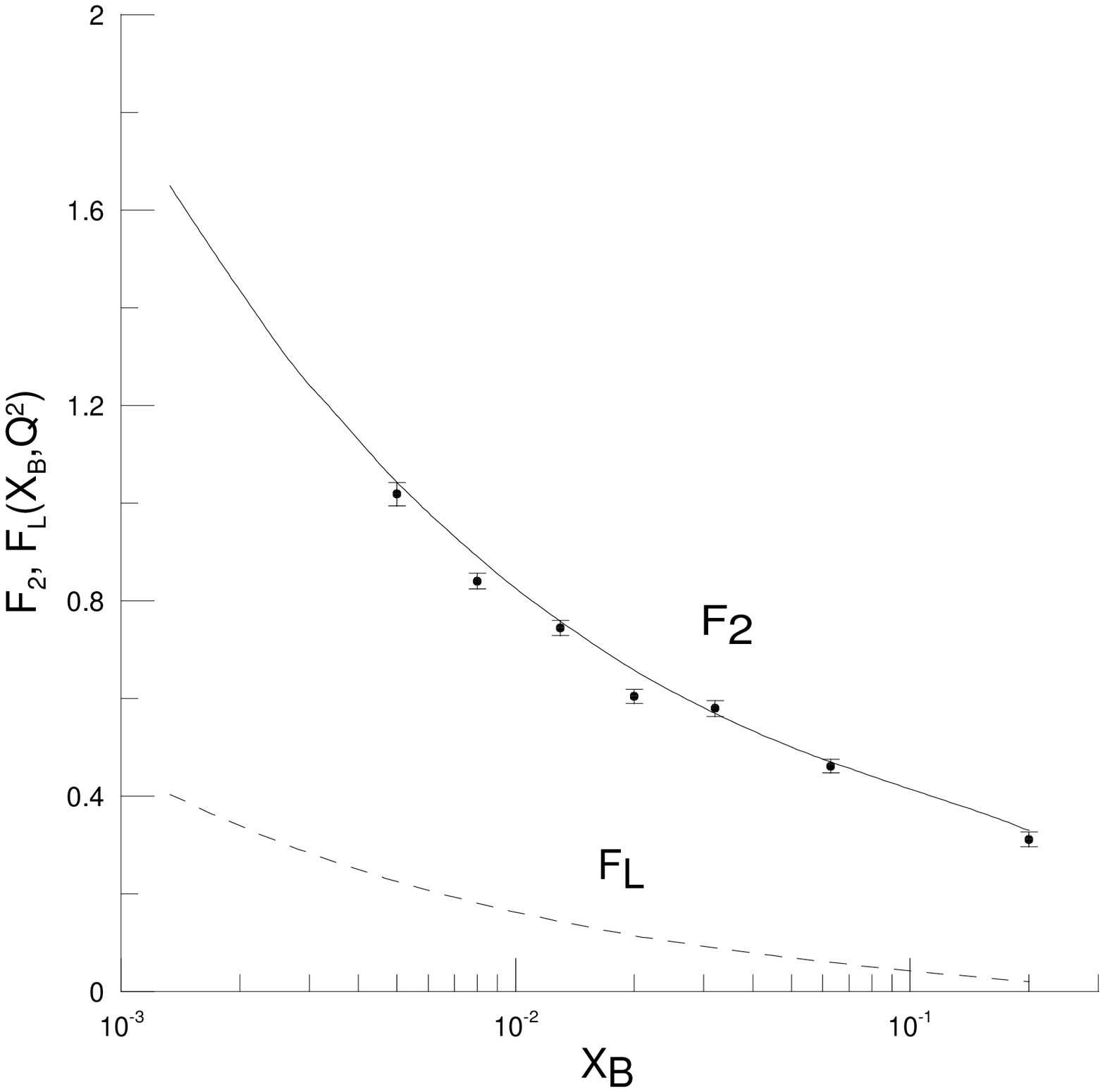}
\end{center}
\caption{$F_2(x_B,Q^2)$ and $F_L(x_B,Q^2)$ at $Q^2=120$ GeV$^2$. The
data are from H1 Collaboration \cite{H1DIS}. \label{fig:f120}}
\end{figure}

Using the formulas (\ref{eq:f2my}) and (\ref{eq:flmy}) we have
calculated DIS structure functions $F_2$ and $F_L$ at the different
values of the photon virtuality $Q^2$, versus the variable $x_B$.
The results are presented in the Figs.~(\ref{fig:f12}) --
(\ref{fig:f120}) in comparison with the H1 Collaboration data
\cite{H1DIS}. We see that the agreement is well. There is no place
for the NLO contribution from the electron -- Reggeized gluon
scattering. At this point our result strongly disagrees with the
result obtained earlier in the Ref.~\cite{WattMartinRyskin}. The
reason of this disagreement is a very approximate choice of the
effective vertex $C_{\gamma Q}^q$ in the
Ref.~\cite{WattMartinRyskin}. They simply take it as in the
collinear approximation, omitting all terms proportional to the
initial quark transverse momentum ${\bf q}_T$. Of course, this
vertex is not gauge invariant and it leads to loss of the of-shell
amplitude correct dependence on the initial quark transverse
momentum. Note, the KMR unintegrated distribution functions are
obtained from conventional collinear distribution functions, and in
the case of quark distributions they include both the sea
quark-originated part (it is large at the lower $x_B$) and the
valence quark-originated part (it is small at the lower $x_B$). The
gluons from a proton contribute effectively in a DIS cross section
via the evolution equations generating a large quark sea at the low
$x_B$.

\boldmath
\section{Prompt photon production at the Tevatron}
\unboldmath \label{sec:foure}

\begin{figure}[ht]
\begin{center}
\includegraphics[width=0.7\textwidth, clip=]{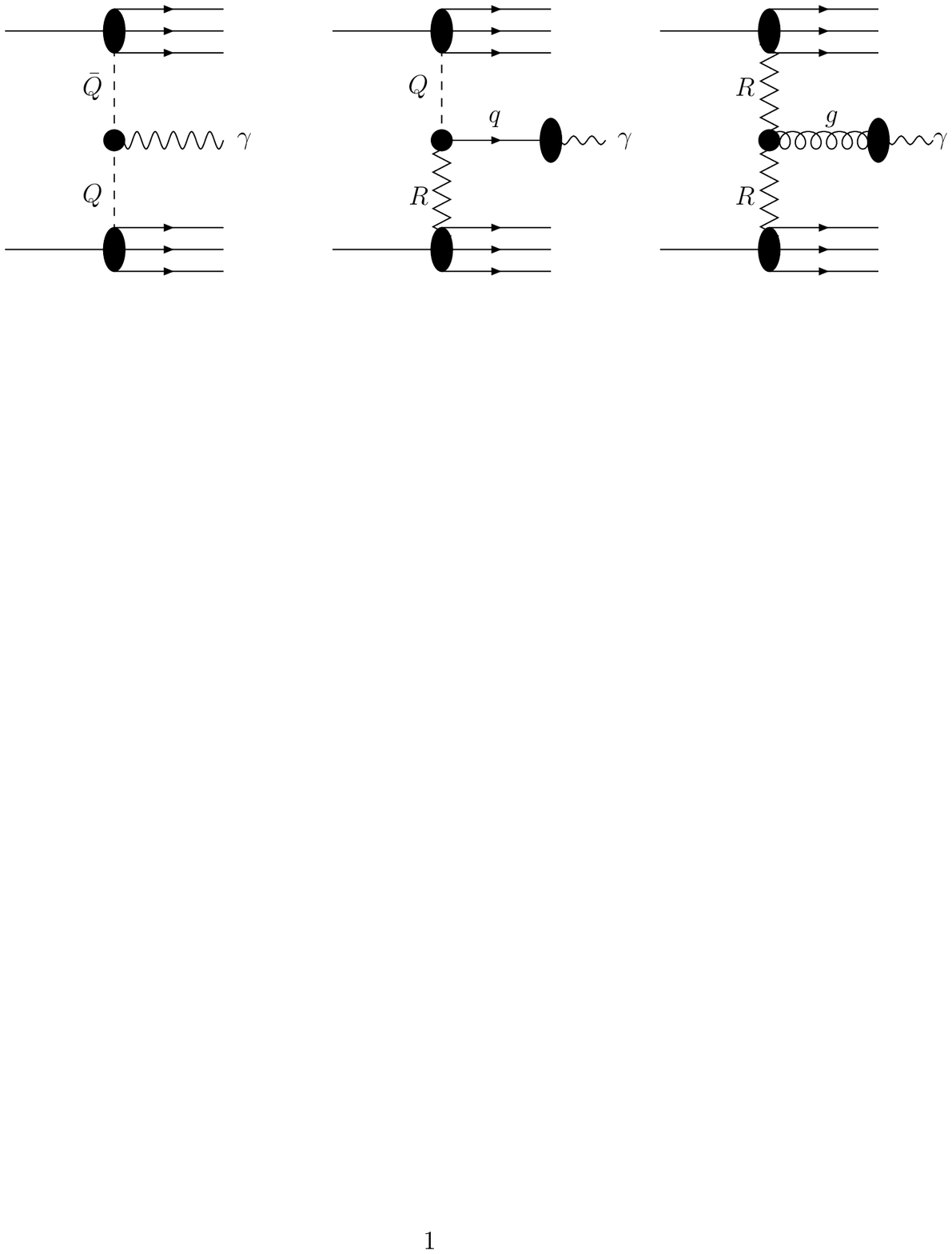}
\end{center}
\caption{The mechanisms of prompt photon hadroproduction in the LO
QMRK approach. \label{fig:tevatron}}
\end{figure}

In this part we consider an inclusive production of isolated photons
at the Fermilab Tevatron Collider, i.e. the processe $p\bar p\to
\gamma X$. In the conventional collinear approximation it is assumed
that at LO prompt photons are produced mainly via quark-gluon
Compton scattering $(qg\to\gamma q$) or quark -- antiquark
annihilation $(q\bar q\to g\gamma)$. In the QMRK approach the LO
contribution comes from the Reggeized quark -- Reggeized antiquark
annihilation via the effective vertices $C_{\bar QQ}^\gamma$
(\ref{eq:QQgamma}). The additional small contributions originate
from fragmentation of produced quarks and gluons into the photon,
which is described by the parton to photon fragmentation functions
$D_{q\to\gamma}(z,\mu^2)$ and $D_{g\to\gamma}(z,\mu^2)$
\cite{DukeOwens}. The Fig.~\ref{fig:tevatron} schematically shows
all relevant contributions to the isolated photon hadroproduction in
the LO QMRK approach.

We start from the factorization formula, which connects the hadron
$d\sigma(p\bar p\to \gamma X)$ cross section with the cross section
of Reggeon's collision $d\hat \sigma(Q\bar Q\to \gamma)$, and which
is presented as follows
\begin{equation}
d\sigma(p\bar p\to \gamma X)=\sum_q\int\frac{dx_1}{x_1}\int
\frac{d^2q_{1T}}{\pi}\int\frac{dx_2}{x_2}\int \frac{d^2q_{2T}}{\pi}
\Phi_{q,\bar q}^p(x_1,t_1,\mu^2)\Phi_{\bar q, q}^{\bar
p}(x_2,t_2,\mu^2) d\hat \sigma(Q\bar Q\to \gamma),
\end{equation}
where $\Phi_{q,\bar q}^{p,\bar p}(x_1,t_1,\mu^2)$ is the Reggeized
quark (anti-quark) unintegrated distribution function in a proton
(anti-proton), $t_{1,2}=-q_{1,2T}^2$, $\mu=p_T$ is taken as the
renormalization, factorization, and fragmentation scale, $p_T$ is
the photon transverse momentum. In the case of photon production via
the fragmentation we need to do a convolution with the relevant
fragmentation function, for example in the following way
\begin{eqnarray}
d\sigma(p\bar p\to \gamma X)&=&\sum_{q,\bar
q}\int\frac{dx_1}{x_1}\int
\frac{d^2q_{1T}}{\pi}\int\frac{dx_2}{x_2}\int
\frac{d^2q_{2T}}{\pi}\int dz \times \nonumber\\
&\times & \Phi_{q,g}^p(x_1,t_1,\mu^2)\Phi_{g, q}^{\bar
p}(x_2,t_2,\mu^2) d\hat \sigma(QR\to q)D_{q\to \gamma}(z,\mu^2),
\end{eqnarray}
where $\Phi_{g}^{p,\bar p}(x_1,t_1,\mu^2)$ is the Reggeized gluon
unintegrated distribution function in a proton (anti-proton). In our
calculations we use simple LO fragmentation functions $D_{q,g\to
\gamma}(z,\mu^2)$, which are taken from Ref.\cite{DukeOwens}.

The  transverse momentum $p_T$ spectra of isolated photon were
studied by CDF\cite{CDF18} Collaboration and D0\cite{D018,D0196}
Collaboration at the energies $\sqrt{S}=1.8$ TeV and $\sqrt{S}=1.96$
TeV. The inclusive prompt photon  production cross section was
measured in the range of $10\leq p_T \leq 300$ GeV, both in the
central region, where the pseudorapidity is $|\eta|<0.9$, and in the
forward region, where one takes on values in the range of
$1.6<|\eta|<2.5$.

The master formula for the differential spectrum of the  directly
produced photon can be presented in the following way:
\begin{equation}
p_T^3\frac{d\sigma}{dp_T}(p\bar p\to \gamma X)=\sum_q\int d\eta \int
d\phi_1\int dt_1\Phi_{q,\bar q}^p(x_1,t_1,\mu^2)\Phi_{\bar q,
q}^{\bar p}(x_2,t_2,\mu^2)\overline{|M(Q\bar Q\to \gamma|^2},
\end{equation}
where $p_0=\frac{1}{2}p_T(e^\eta+e^{-\eta})$ is the photon energy,
$p_z=\frac{1}{2}p_T(e^\eta-e^{-\eta})$ is the photon longitudinal
momentum in respect of the proton beam, $\phi_1$ is the angle
between ${\bf p}_T$ and ${\bf q}_{1T}$, and
$$x_1=\frac{p_0+p_z}{\sqrt{S}}, \quad x_2=\frac{p_0-p_z}{\sqrt{S}},
 \quad t_2=t_1-2p_T\sqrt{t_1}\cos (\phi_1)+p_T^2,\quad p_T^2=x_1x_2S.$$

In the case of the photon production via a quark fragmentation the
photon $p_T-$spectrum  can be presented as follows
\begin{eqnarray}
p_T^3\frac{d\sigma}{dp_T}(p\bar p\to \gamma X)&=&\sum_{q,\bar q}\int
d\eta \int d\phi_1\int dt_1\int dz~
\Phi_{q,g}^p(x_1,t_1,\mu^2)\Phi_{g, q}^{\bar
p}(x_2,t_2,\mu^2)\times\nonumber\\
&\times& z^2 D_{q\to \gamma}(z,\mu^2)\overline{|M(QR\to q|^2},
\end{eqnarray}
where
$$x_1=\frac{p_0+p_z}{\sqrt{S}z}, \quad x_2=\frac{p_0-p_z}{\sqrt{S}z},
\quad t_2=t_1-2\frac{p_T}{z}\sqrt{t_1}\cos (\phi_1)+\frac{p_T^2}{z^2}.$$

\begin{figure}[ht]
\begin{center}
\includegraphics[width=0.45\textwidth, clip=]{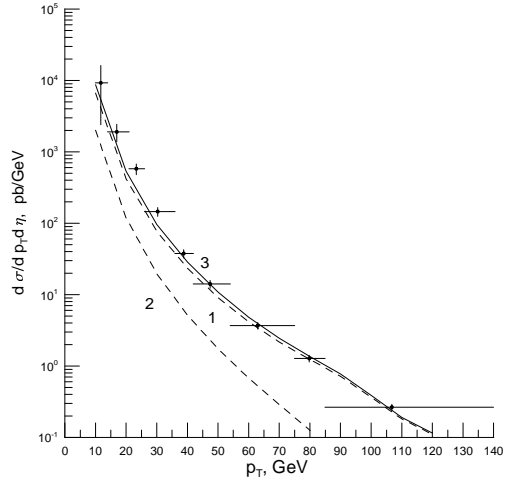}
\end{center}
\caption{The $p_T-$spectrum of prompt photon at $\sqrt{S}=1.8$ TeV
and $|\eta|<0.9$. The curve 1 is the direct production, 2 is the
fragmentation production, 3 is their sum. The data are from D0
Collaboration \cite{D018}.\label{fig:tev181}}
\end{figure}

\begin{figure}[ht]
\begin{center}
\includegraphics[width=0.45\textwidth, clip=]{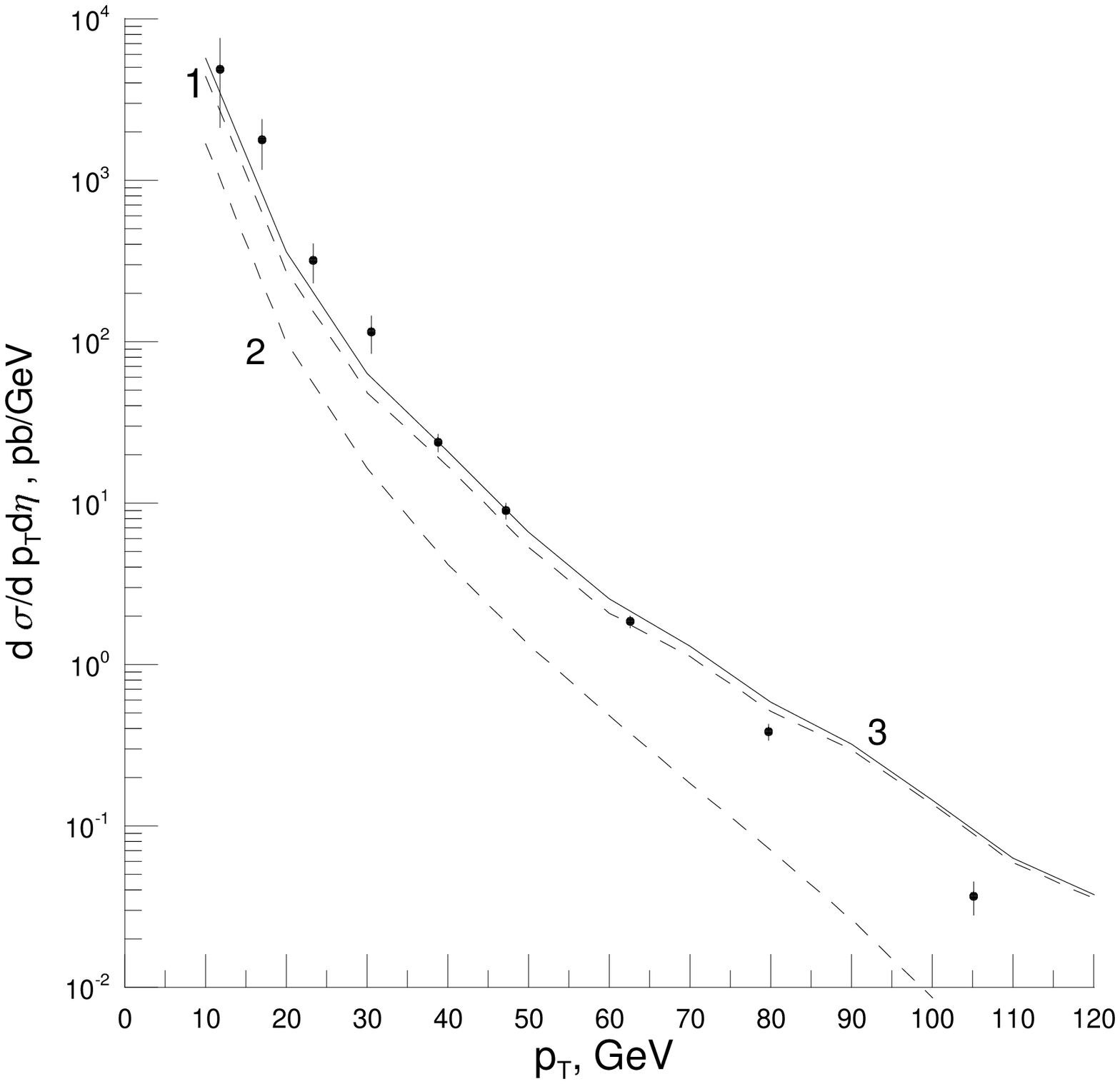}
\end{center}
\caption{The $p_T-$spectrum of prompt photon at $\sqrt{S}=1.8$ TeV
and $1.6<|\eta|<2.5$. The curves are the same as in
Fig.~\ref{fig:tev181}. The data are from D0 Collaboration
\cite{D018}. \label{fig:tev182}}
\end{figure}

\begin{figure}[ht]
\begin{center}
\includegraphics[width=0.45\textwidth, clip=]{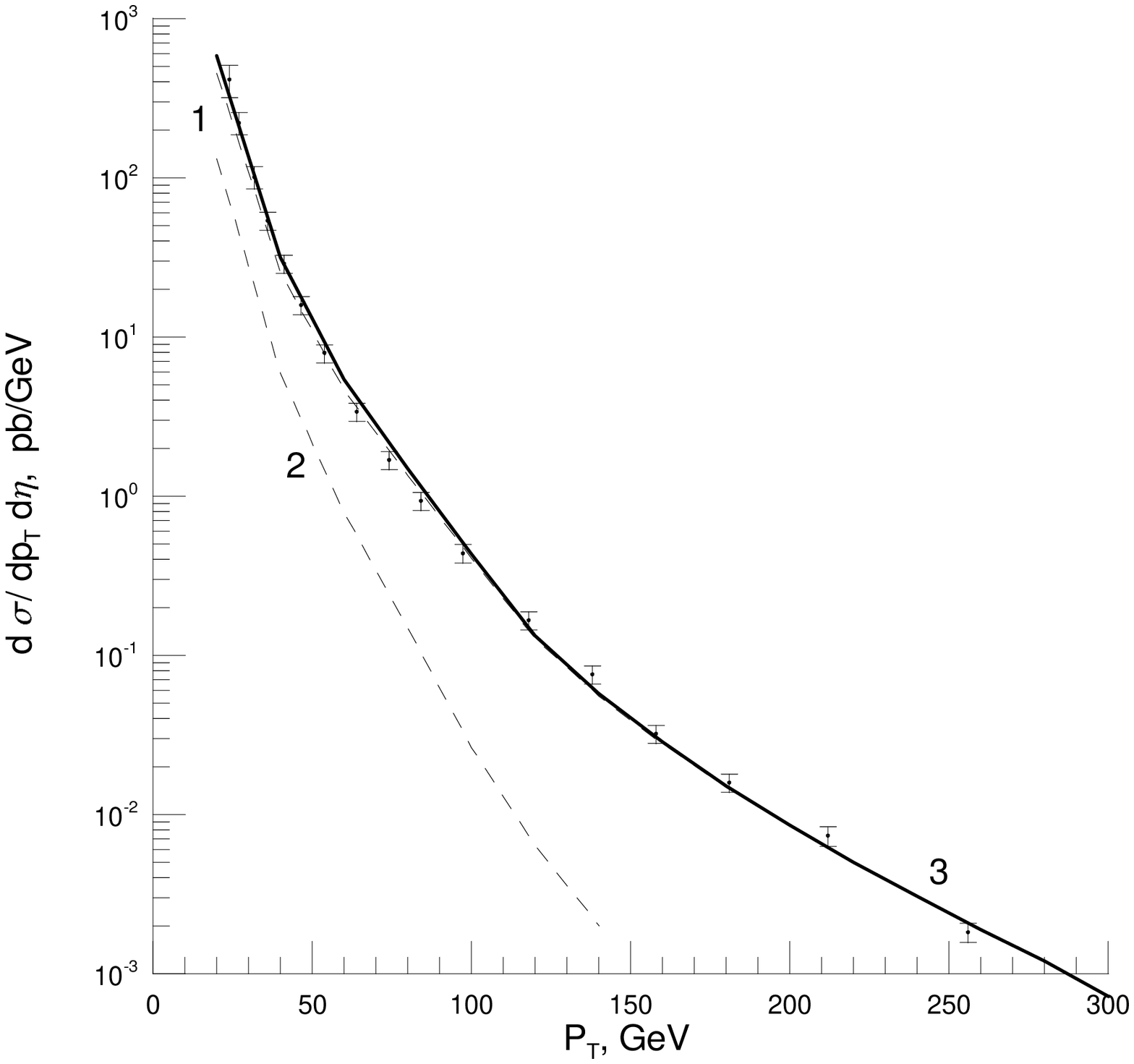}
\end{center}
\caption{The $p_T-$spectrum of prompt photon at $\sqrt{S}=1.96$ TeV
and $|\eta|<0.9$ . The curves are the same as in
Fig.~\ref{fig:tev181}. The data are from D0 Collaboration
\cite{D0196}. \label{fig:tev196}}
\end{figure}

In the Figures (\ref{fig:tev181}), (\ref{fig:tev182}) and
(\ref{fig:tev196}) we present the results of our calculation in
comparison with the D0 Collaboration data \cite{D018,D0196} at the
energy $\sqrt{S}=1.8$ Tev and $\sqrt{S}=1.96$ TeV, correspondingly.
We see that the agreement between the data and the theoretical
calculation is well, especially in the central region of a
pseudorapidity $|\eta|<0.9$. The contribution of the fragmentation
production mechanism is very small at the $\sqrt{S}=1.8$ TeV and it
is negligible at the energy $\sqrt{S}=1.96$ TeV. It is a very
interesting fact, that at the energy $\sqrt{S}=1.96$ TeV we describe
the data up to $p_T \simeq 300$ GeV. That directly demonstrates
valid behavior and correct normalization of the quark unintegrated
structure function $\Phi_{q}^{p}(x,t,\mu^2)$, which is obtained
using KMR prescription \cite{KMR}, in the wide region of parameters
$x$ and $t$.

\boldmath
\section{Discussion}
\unboldmath \label{sec:five}

In this section let us to compare our results with the previous
studies of DIS and prompt photon production in the
$k_T-$factorization approach, which were published  in the Refs.
\cite{WattMartinRyskin, KMRgamma, ZotovLipatovGamma, SzczurekGamma}
recently.

The electron -- quark DIS in the $k_T-$factorization approach was
considered in the Ref.~\cite{WattMartinRyskin}. It was assumed that
in DIS the transverse momentum of the off-shell quark much smaller
than the photon virtuality ${\bf k}_T^2\ll Q^2$, and the
$k_T-$dependence in the numerator of the relevant amplitude  was
ignored. In other words, in this approximation the conventional
gamma-quark vertex was used, and gauge dependence of the off-shell
amplitude was lost. It was obtained for the DIS structure functions
\cite{WattMartinRyskin}:
\begin{eqnarray}
F_2(x_B,Q^2)&=&\sum_{q,\bar q}e_q^2\int
dt_2\left(\frac{x_B}{x_2}\right)
\Phi_{q,\bar q}(x_2,t_2,\mu^2),\label{eq:f2kmr}\\
F_L(x_B,Q^2)&=&0.\label{eq:fLkmr}
\end{eqnarray}
The differences between  our formulas (\ref{eq:f2my})--
(\ref{eq:flmy}) and ones (\ref{eq:f2kmr}) -- (\ref{eq:fLkmr}),
demonstrate that gauge invariance condition for amplitudes controls
the correct $k_T-$dependence of the DIS structure functions.  The
formulas (\ref{eq:f2kmr}) and (\ref{eq:fLkmr}) don't describe the
experimental data at the LO and one needs the NLO contribution
coming from the photon-gluon fusion subprocesses. The calculation of
a cross section within the QMRK approach  for the process under
consideration
\begin{equation}
\gamma^\star + R \to q + \bar q,
\end{equation}
where $q$ is the massless quark, needs especial analysis with
extraction of different singularities, see
Ref.~\cite{FadinIvanovKotsky} for example. The calculation of DIS
structure functions in the QMRK approach at the NLO has not been
made up today. As it was shown above, even the LO approximation with
the Reggeized quark describes data well, and this NLO contribution
should be small.

The calculations of the inclusive prompt photon production at the
Tevatron were performed using off-shell amplitudes involving initial
quark in the Refs.~\cite{KMRgamma, ZotovLipatovGamma,
SzczurekGamma}. In all these papers the LO contribution coming from
the annihilation of a Reggeized quark and a Reggeized anti-quark
($Q+\bar Q\to \gamma$) was ignored and  a Compton process of the
off-shell quark -- off-shell gluon scattering $g^\star + q^\star \to
\gamma + q$ was considered as the LO process. We can't explain this
approximation. Evidently, if we work with off-shell initial quarks
we need to take into account the process $q^\star+\bar q^\star \to
\gamma$ as the LO approximation. We especially used here different
denotations for the Reggeized particles ($Q,R$) and off-shell
particles ($q^\star, g^\star$). It means that in the
Refs.~\cite{KMRgamma, ZotovLipatovGamma, SzczurekGamma} the
conventional QCD vertices are used and the authors have obtained the
gauge uninvariant amplitudes. The inclusion of the $k_T$-effects in
the kinematics only without the using  of the correct vertices for
Reggeized particle interactions, can be considered as
phenomenological trick, but such inclusion has not predictive power.
The Compton scattering of Reggeized gluons on Reggeized quarks,
\begin{equation}
R + Q \to \gamma + q,
\end{equation}
is a NLO QMRK process, when it is used  to calculate inclusive
photon production one needs to solve the same problems as for other
NLO Reggeon-Reggeon to Particle-Particle processes in the QMRK
approach. First of all, it is necessary to obtain the relevant
effective vertices. This task have not been  solved yet. Of course,
the relevant process is LO if we study the associated photon plus
jet, both with the larger transverse momenta, production.

 \boldmath
\section{Conclusion}
\unboldmath \label{sec:six}

We have shown that the quark Reggeization hypothesis is a very
powerful tool in the high energy phenomenology for the hard
processes involving quark exchanges. It is shown that it is possible
to describe data at the LO QMRK approach for the DIS structure
functions $F_2$ and $F_L$ of lepton -- proton scattering, and prompt
photon spectra for the inclusive production in the $p\bar p$
collisions at high energies. The scheme suggested in this paper is
principally new in comparison with both the conventional collinear
approximation and the $k_T-$factorization approach, which was
developed earlier using Reggeized gluons only.

\section{Acknowledgements}
We thank  L.~Lipatov for the censorious  remarks and useful
discussion of the questions under consideration in this paper. We
also thank D.~Vasin and A.~Shipilova for help in numerical
calculations.

\newpage
%\newpage

\newpage

\newpage

\newpage

\newpage

\begin{thebibliography}{99}
\bibitem{GribovLevinRyskin}
L.~V.~Gribov, E.~M.~Levin, and M.~G.~Ryskin, Phys.\ Rept.\
\textbf{100}, 1 (1983).

\bibitem{CollinsEllis}
J.~C.~Collins and R.~K.~Ellis, Nucl.\ Phys.\ \textbf{B360}, 3
(1991).

\bibitem{CataniCH}
S.~Catani, M.~Ciafoloni, and F.~Hautmann, Nucl.\ Phys.\
\textbf{B366}, 135 (1991).

\bibitem{FadinLipatov96}
V.~S.~Fadin and L.~N.~Lipatov, Nucl.\ Phys.\ \textbf{B477}, 767
(1996); Nucl.\ Phys.\ \textbf{B406}, 259 (1993).

\bibitem{KniehlSaleevVasin}B.~A.~Kniehl, V.~A.~Saleev, and D.~V.~Vasin,
Phys.\ Rev.\ D \textbf{73}, 074022 (2006); Phys.\ Rev.\ D
\textbf{74}, 014024 (2006); V.~A.~Saleev and D.~V.~Vasin, Phys.\
Rev.\ D \textbf{68}, 114013 (2003); V.~A.~Saleev, Phys.\ Rev.\ D
\textbf{65}, 054041 (2002).



\bibitem{Lipatov95} L.~N.~Lipatov, Nucl.\ Phys.\ \textbf{B452}, 369 (1995).

\bibitem{NLO}
V.~S.~Fadin, M.~I.~Kotsky, and L.~N.~Lipatov, Phys.\ Lett.\ B
\textbf{415}, 97 (1997); D.~Ostrovsky, Phys.\ Rev.\ D \textbf{62},
054028 (2000); V.~S.~Fadin, M.~G.~Kozlov, and A.~V.~Reznichenko.
Report No. DESY 03-025 (2003); J.~Bartels, A.~S.~Vera, and
F.~Schwennsen, JHEP 0611: 051 (2006).

\bibitem{FadinSherman} V.~S.~Fadin and V.~E.~Sherman, JETP Lett. \textbf{23},
548 (1976); JETP \textbf{45}, 861 (1977).

\bibitem{D018} D0 Collaboration,
B.~Abbott \emph{et al.}, Phys.\ Rev.\ Lett.\  \textbf{84}, 2786
(2000).

\bibitem{CDF18} CDF Collaboration, D.~Acosta \emph{et al.}, Phys.\ Rev.\ D \textbf{70},
074008 (2004).

\bibitem{D0196} D0 Collaboration,
V.~M.~Abazov \emph{et al.}, Phys.\ Lett.\ B \textbf{639}, 151
(2006).

%\bibitem{ZEUSgamma} ZEUS Collaboration, J.~Breitweg \emph{et al.},
%Phys.\ Lett.\ B \textbf{472}, 175 (2000).
%
%\bibitem{H1gamma} H1 Collaboration, A.~Aktas \emph{et al.}, Eur.\ Phys.\ J.\, C  \textbf{38}, 437 (2004).
%
\bibitem{KMR} M.~A.~Kimber, A.~D.~Martin, and M.~G.~Ryskin,
Phys.\ Rev.\ D \textbf{63}, 114027 (2001).

\bibitem{KMRgamma} M.~A.~Kimber, A.~D.~Martin, and M.~G.~Ryskin,
Eur.\ Phys.\ J.\ C \textbf{12}, 655, (2000).

\bibitem{ZotovLipatovGamma} A.~V.~Lipatov and N.~P.~Zotov, Phys.\
Rev.\ D \textbf{72}, 054002 (2005).

\bibitem{SzczurekGamma} T.~Pietrycki and A.~Szczurek, Phys.\ Rev.\ D
\textbf{75}, 014023 (2007).

\bibitem{BFKL}
E.~A.~Kuraev, L.~N.~Lipatov, and V.~S.~Fadin, Sov.\ Phys.\ JETP
\textbf{44}, 443 (1976) [Zh.\ Eksp.\ Teor.\ Fiz.\  \textbf{71}, 840
(1976)]; I.~I.~Balitsky and L.~N.~Lipatov, Sov.\ J.\ Nucl.\ Phys.\
\textbf{28}, 822 (1978) [Yad.\ Fiz.\ \textbf{28}, 1597 (1978)].

\bibitem{KTAntonov}
E.~N.~Antonov, L.~N.~Lipatov, E.~A.~Kuraev, and I.~O.~Cherednikov,
Nucl.\ Phys.\ B \textbf{721}, 111 (2005).

\bibitem{LipatovFadin89} L.~N.~Lipatov and V.~S.~Fadin, JETP Lett. \textbf{49}, 352
(1989); Sov.\ J.\ Nucl.\ Phys.\ \textbf{50}, 712 (1989).

\bibitem{GellMann} M.~Gell-Mann, M.~L.~Goldberger, F.~E.~Low,
E.~Marx, and F.~Zachariasen, Phys.\ Rev.\ \textbf{133}, 145B (1964).


\bibitem{FadinFiore} V.~S.~Fadin and R.~Fiore, Phys.\
Rev.\ D \textbf{64}, 114012 (2001).

\bibitem{FadinBogdan} A.~V.~Bogdan and V.~S.~Fadin, Nucl.\
Phys.\ B \textbf{740}, 36 (2006).


\bibitem{WattMartinRyskin} G.~Watt, A.~D.~Martin and M.~G.~Ryskin,
Eur.\ Phys.\ J.\ C \textbf{31}, 73 (2003); Phys.\ Rev.\ D
\textbf{70}, 014012 (2004).

\bibitem{H1DIS} H1 Collaboration, J.~C.~Adloff et al., Eur.\ Phys.\
J.\ C \textbf{21}, 33 (2001).

\bibitem{DukeOwens} D.~W.~Duke and J.~F.~Owens, Phys.\ Rev.\ D \textbf{26},
1600 (1982); J.~F.~Owens, Rev.\ Mod.\ Phys.\ \textbf{59}, 465
(1987).




\bibitem{FadinIvanovKotsky} V.~S.~Fadin, D.~Yu.~Ivanov and M.~I.~Kotsky,
Nucl.\ Phys.\ B \textbf{658}, 156 (2003).

\end{thebibliography}
\end{document}